\documentclass[
aps,
prb,
showpacs,
floatfix,
reprint,
twoside,
superscriptaddress
]{revtex4-1}
\usepackage{amsmath}
\usepackage[T1]{fontenc}
\usepackage{fourier}
\usepackage{graphicx}
\usepackage[colorlinks=true,
allcolors=blue,
dvipdfm=true]{hyperref}

\begin{document}

\title{Topological Defects in Systems with Two Competing Order Parameters:\\
Application to Superconductors with Charge- and Spin-Density Waves}

\author{Andreas Moor}
\affiliation{Theoretische Physik III, Ruhr-Universit\"{a}t Bochum, D-44780 Bochum, Germany}
\author{Anatoly F.~Volkov}
\affiliation{Theoretische Physik III, Ruhr-Universit\"{a}t Bochum, D-44780 Bochum, Germany}
\author{Konstantin B.~Efetov}
\affiliation{Theoretische Physik III, Ruhr-Universit\"{a}t Bochum, D-44780 Bochum, Germany}
\affiliation{National University of Science and Technology ``MISiS'', Moscow, 119049, Russia}

\begin{abstract}
On the basis of coupled Ginzburg--Landau equations we study nonhomogeneous states in systems with two order parameters~(OP). Superconductors with superconducting OP~$\Delta$, and charge- or spin-density wave (CDW or SDW) with amplitude~$W$ are examples of such systems. When one of OP, say~$\Delta$, has a form of a topological defect, like, e.g., vortex or domain wall between the domains with the phases~$0$ and~$\pi$, the other OP~$W$ is determined by the Gross--Pitaevskii equation and is localized at the center of the defect. We consider in detail the domain wall defect for~$\Delta$ and show that the shape of the associated solution for~$W$ depends on temperature and doping (or on the curvature of the Fermi surface)~$\mu$. It turns out that, provided temperature or doping level are close to some discrete values~$T_{n}$ and~$\mu_{n}$, the spacial dependence of the function~$W(x)$ is determined by the form of the eigenfunctions of the linearized Gross--Pitaevskii equation. The spacial dependence of~$W_{0}$ corresponding to the ground state has the form of a soliton, while other possible solutions~$W_{n}(x)$ have nodes. Inverse situation~when~$W(x)$ has the form of a topological defect and~$\Delta(x)$ is localized at the center of this defect is also possible. In particular, we predict a surface or interfacial superconductivity in a system where a superconductor is in contact with a material that suppresses~$W$. This superconductivity should have rather unusual temperature dependence existing only in certain intervals of temperature. Possible experimental realizations of such non-homogeneous states of OPs are discussed.
\end{abstract}

\date{\today}

\maketitle

\section{Introduction}

Materials with two order parameters~(OP) have long been known. For example, superconductivity in stoichiometric ternary compounds~ErRh$_{4}$B$_{4}$ and~M$_{x}$Mo$_{6}$S$_{8}$ (with~M meaning~Ho,~Dy, or~Er and ${x=1}$ or~$1.2$) coexists with a helical magnetic order.~\cite{Bulaev85,*Bulaev85a} In the last decades interest in the systems with two OPs has increased drastically in connection with the discovery of high\nobreakdash-$T_{\text{c}}$ superconductors (see, e.g.,~\onlinecite{Kivelson03,Vojta09,SachdevCDW,Efetov13,*SachdevCDWa,Meier14,Chubukov14,Fradkin14}). Very recently, it has been experimentally established that, in cuprates, the superconducting OP~$\Delta $ coexists with a state with a charge modulation (see recent papers Refs.~\onlinecite{CDWexp1,CDWexp2,CDWexp3,CDWexp4,CDWexp5,CDWexp6,CDWexp7,CDWexp8,CDWexp9,CDWexp10} and references therein).

In another class of recently discovered superconductors---the so-called Fe\nobreakdash-based pnictides---superconductivity may coexist with a spin density wave~(SDW) (for a review see Refs.~\onlinecite{Hirschfeld_Korshunov_Mazin_2011,SDWrev}).

Coexistence of OPs of different types results in several interesting phenomena. One can mention the enhancement of the London penetration depth\cite{ExperLondon,ChubukovLondon,SachdevLondon} or a peak in the specific heat jump\cite{ExperHeat,Levchenko14} at the doping level at which the SDW is formed, the peculiar dynamics of the OPs~(see recent papers Refs.~\onlinecite{Sachdev14,MVVE14} and references therein).

Nonhomogeneous states in systems with two OPs are also very interesting and unusual. For example, a CDW arises in the center of vortices in cuprates.\cite{CDWvortex} Nonhomogeneous states in superconductors may arise even in the absence of magnetic field. For example, Fulde-Ferrel-Larkin-Ovchinnikov~(FFLO) states may appear in superconductors in the presence of an exchange field\cite{Larkin_Ovchinnikov_1965,*Fulde_Ferrell_1964} and the so called amplitude solitons can be energetically favorable in conductors with CDW or SDW. The latter have been predicted in Refs.~\onlinecite{Heeger79,Braz79,*Braz79a,*Braz79b} and observed in Ref.~\onlinecite{Monceau} in systems with a single OP---in quasi-one-dimensional conductors with a CDW.

Amplitude solitons in systems with a CDW mean that the amplitude~$W(x)$ of the CDW drops to zero at some point, and a local energy level~$\epsilon_{0}$ arises in the system.\cite{HeegerRev} For example, if ${W = W_{\infty} \tanh (x / \sqrt{2} \xi_{w})}$, the phase~$\chi$ of~$W$ changes from ${\chi = 0}$ at ${x = \infty}$ to ${\chi = \pi}$ at ${x = -\infty}$. The energy level~$\epsilon_{0}$ may vary in time when a sufficiently high current passes through the system. In this case, the stationary state is unstable and one deals with a dynamical amplitude soliton with~$\epsilon_{0}(t)$.\cite{AVK86} Another case of a non-stationary (moving) soliton was studied in a recent work.\cite{Galitski14,Artemenko_2003} In both cases, the structure of the amplitude soliton can be found analytically from a solution of microscopic equations.\cite{AVK86,Galitski14} It is relevant to mention the stripes in high\nobreakdash-$T_{\text{c}}$ superconductors\cite{zaanen,machida,poilblanc,white_1998,tranquada} that are higher dimensional relatives of the solutions of Refs.~\onlinecite{Heeger79,Braz79,*Braz79a,*Braz79b}.

Fulde-Ferrel-Larkin-Ovchinnikov states in superconductors coexisting with other OPs, such as CDW or SDW, were studied recently in several works.\cite{Agterberg08,GorkovTeitel} In particular, it has been pointed out in Ref.~\onlinecite{GorkovTeitel} that the states similar to the~FFLO state are possible in superconductors competing with CDW or SDW. Note that another type of nonhomogeneous states in systems with two OPs has been studied in Refs.~\onlinecite{Nussinov_2004,Wu_Schmalian_Kotliar_Wolynes_2004,Nussinov_Vekhter_Balatsky_2009} using the so-called Brazoskii-type model.\cite{Brazovskii_1975} In particular, it has been shown that a ``glassy'' phase may arise in these systems.

However, the FFLO-like state may arise, e.g., in Fe\nobreakdash-based pnictides, only at low enough temperatures~$T$ when the dependence of~$W$ on the curvature~$\mu$ is a multivalued function\cite{Chubukov10,*Chubukov10a,Schmalian10,Moor11} which is in a full analogy with superconductors with an exchange field~$h$. In the latter case, the dependence~$\Delta(h)$ at low~$T$ is a multivalued function in a certain interval of~$h$.\cite{Larkin_Ovchinnikov_1965,*Fulde_Ferrell_1964} Spatial dependence of~$\Delta$ and~$W$ can be described by a generalized Eilenberger equation complemented by self-consistency relations\cite{Moor11} but their analytical study is not simple.

In this Paper, we analyze nonuniform states of the OPs on the basis of the Ginzburg--Landau equations that are considerably more transparent than the Eilenberger equation. We are interested in nonuniform states corresponding to topological defects. We concentrate here on one-dimensional structures and consider the dependence only on one coordinate~$x$. This means that we consider a situation when the superconducting OP~$\Delta$ changes its phase from~$0$ to~$\pi$ across this defect, while the amplitude of the CDW (or SDW)~$W(x)$ is localized at the center of the defect decaying to zero away from this point. Of course, the opposite situation is also possible, i.e., when the function~$W(x)$ changes sign having opposite values at~$-\infty$ and~$\infty$ and~$\Delta(x)$ is a localized function.

Using a system of coupled Ginzburg--Landau equations we show that, while the superconducting OP may vary in space as ${\Delta(x) = \Delta_{\infty} \tanh (x / \sqrt{2} \xi_{s})}$, the form of the amplitude~$W(x)$ depends on temperature or doping. It is described by the Gross--Pitaevskii equation resulting in a peculiar quantization of the solutions for~$W(x)$. The function~$W_{0}(x)$ corresponding to the ground state has the form of a soliton whereas the functions~$W_{n}(x)$ corresponding to excited states have nodes.

In the opposite case, when a solution ${W(x) = W_{\infty} \tanh (x / \sqrt{2} \xi_{w})}$ is brought about, the superconducting OP~$\Delta$ is localized near the point~${x = 0}$ and changes its form with variation of temperature in a rather unusual way.

The nonuniform states considered in this Paper may arise in the bulk, near the surface or in heterostructures consisting of materials with two OPs. In the latter cases such states appear necessarily because of boundary or matching conditions imposed on the OPs. Note that the interesting and novel phenomena arising at the surface or in
heterostructures are already known. A new type of superconductivity (triplet odd-frequency) in superconductor/ferromagnet bilayer in the ferromagnet\cite{Bergeret_Volkov_Efetov_2001,*Bergeret_Volkov_Efetov_2005} and the appearance of bound edge states with possible formation of Majorana fermions  at the surface of superconductors\cite{Hasan_Kane_2010,*Tanaka_Sato_Nagaosa_2012} are remarkable examples of these phenomena.

\section{Free Energy and Ginzburg--Landau Equations}

We consider a model which is described by the G\nobreakdash--L~equations. As has been shown in Ref.~\onlinecite{MVVE14}, it is applicable to quasi-one-dimensional superconductors with a CDW and to two-band superconductors with an SDW. The latter model has been developed in detail in Refs.~\onlinecite{Chubukov10,*Chubukov10a,Schmalian10} for Fe\nobreakdash-based pnictides. After certain modification, this model can be applied also to cuprates.\cite{MVVE14} On its basis, one can derive G\nobreakdash--L~equations for the OPs~$\Delta$ and~$W$. We neglect space variations of the phases of~$\Delta$ and~$W$ and consider these OPs as real quantities.

To make the physical meaning of the coefficients in the G\nobreakdash--L~expansion more transparent, we write the G\nobreakdash--L~equations first for the case of superconductors with a CDW (or~SDW). In the notation of Refs.~\onlinecite{Chubukov10,*Chubukov10a,Schmalian10}, these equations have the form
\begin{align}
-\xi_{s}^{2} \nabla^{2} \Delta + \Delta \big[ W^{2} s_{2m} + \Delta^{2} s_{3} - \ln(T_{s}/T) \big] &= 0 \,, \label{1} \\
-\xi_{w}^{2} \nabla^{2} W + W \big[ \langle 2 \mu^{2} s_{1m} \rangle + W^{2} s_{3m} + \Delta^{2} s_{2m} - \ln(T_{w}/T) \big] &= 0 \,, \label{1'}
\end{align}
where~$\xi_{s,w}^{2}$ are the coherence lengths (at low temperatures) for~$\Delta$ and~$W$, respectively, and~$T_{s,w}$ are, respectively, the critical temperatures for the transition into the pure superconducting state or into a state with a CDW or an SDW only. In other words,~$T_{w}$ is the critical temperature for the transition into the charge-ordered state in absence of~$\Delta$ and~$\mu$, while~$T_{s}$ is the superconducting transition temperature in absence of~$W$. The angle brackets mean the angle averaging (in Fe\nobreakdash-based pnictides) or integration along the sheets of the Fermi surfaces in quasi-one-dimensional superconductors. The functions~$s_{1m}$,~$s_{2m}$, etc.,~are functions of the normalized curvature (see Appendix) ${m=\mu /(\pi T_{s})}$ and ${\mu = \mu_{0} + \mu_{\varphi} \cos \big[(p_{y}^{2} + p_{z}^{2})^{1/2} a \big]}$ is a curvature in quasi-one-dimensional superconductors with a doping-dependent value of~$\mu_{0}$. It is assumed that the Fermi surface of these superconductors consists of two slightly curved sheets which are perpendicular to the~$x$~axis.\cite{MVVE14} In the case of Fe\nobreakdash-based pnictides, ${\mu = \mu_{0} + \mu_{\varphi} \cos (2\varphi)}$ is a quantity that describes an elliptic (${\mu_{\varphi} \neq 0}$) and circular (${\mu_{\varphi} = 0}$) Fermi surfaces of electron and hole bands.\cite{Chubukov10,*Chubukov10a,Schmalian10} All quantities---$\Delta$,~$W$ and~$\mu $---are measured in units of~$\pi T_{s}$. The expressions for the coefficients in the G\nobreakdash--L~expansion with account for impurity scattering have been calculated in Ref.~\onlinecite{Syzranov14} (see also Ref.~\onlinecite{Vavilov11a}).

Replacing the derivative ${\nabla \to \nabla -\mathrm{i} 2 \pi A / \Phi_{0}}$, one can use Eqs.~\eqref{1} and~\eqref{1'} to describe vortices in superconductors with a CDW~\cite{Efetov13a}, where~$\Phi_{0}$ is the magnetic flux quantum and~$A$--the modulus of the vector potential of a magnetic field.

As it is seen from Eq.~\eqref{1'}, the critical temperature~$T_{w}$ depends on doping, i.e., on the parameter~$\mu$. We choose this parameter ${\mu =
\mu_{c}}$ in such a way that ${T_{w}(\mu_{c}) = T_{s}}$. This means that at ${T = T_{s}}$, the quantities ${\Delta = W = 0}$, and, thus,~$\mu_{c}$ obeys
the equation
\begin{equation}
\langle 2 \mu_{c}^{2} s_{1m}(\mu_{c}) \rangle = \ln ( T_{w}/T_{s} ) \equiv \ln r \,, \label{2}
\end{equation}
where ${r = T_{w}/T_{s}}$ and~$\mu_{c}$ is a function of two parameters, i.e.,~${\mu_{c} = \mu_{c}(\mu_{0}, \mu_{\varphi}})$.

Then, we expand the function~$s_{1m}(\mu ,T)$ in the deviations ${\delta [\mu^{2}] = \mu^{2} - \mu_{c}^{2}}$ and ${\delta T = T_{s} - T}$, thus obtaining ${s_{1m}(\mu, T) = s_{1m}(\mu_{c}, T_{s}) + \beta_{1} \delta T + \langle \beta_{2} \delta [\mu^{2}] \rangle}$, and use Eq.~(\ref{2}) to obtain equations in a general standard form (assuming that all the
functions depend only on one coordinate~$x$),
\begin{align}
-\xi_{s}^{2} \Delta^{\prime \prime} + \Delta \big[ -a_{s} + b_{s} \Delta^{2} + \gamma W^{2} \big] &= 0 \,, \label{3} \\
-\xi_{w}^{2} W^{\prime \prime} + W \big[ -a_{w} + b_{w} W^{2} + \gamma \Delta^{2} \big] &= 0 \,, \label{3'}
\end{align}
with~$\Delta^{\prime}$ and~$W^{\prime}$ as well as~$\Delta^{\prime \prime}$ and~$W^{\prime \prime}$ denoting the first and second derivatives with respect to~$x$, respectively. These equations determine extrema of the free energy functional
\begin{align}
\mathcal{F} = \frac{1}{2} \int \mathrm{d} x \big\{& \xi_{s}^{2} \Delta^{\prime 2} - a_{s} \Delta^{2} + \frac{b_{s}}{2} \Delta^{4} + \gamma \Delta^{2} W^{2} \notag \\
+& \xi_{w}^{2} W^{\prime 2} - a_{w} W^{2} + \frac{b_{w}}{2} W^{4} \big\}  \label{4}
\end{align}
with respect to~$\Delta$ and~$W$, and the corresponding coefficients of the G\nobreakdash--L~expansion are related to variables in Eqs.~(\ref{1}) and~(\ref{1'}) via ${a_{s} = \eta}$, ${b_{s} = s_{3} \simeq 1.05}$, ${a_{w} = \eta (1 - \beta_{1}) - \langle \beta_{2} \delta [\mu^{2}] \rangle }$, ${b_{w} = s_{3m}}$, ${\gamma = s_{2m}}$, where ${\eta = 1 - T/T_{s}}$. The expressions for~$\beta _{1,2}$ are
given in the Appendix.

The coupled G\nobreakdash--L Eqs.~(\ref{3}) and~(\ref{3'}) are, of course, not new and have been used long time ago in, e.g., Ref.~\cite{Efetov81,*Efetov81a} for studying competition between superconductivity and CDW in the presence of disorder or commensurability.

\section{Soliton-like solutions at quantized temperatures and doping}

Our aim now is to find new non-trivial inhomogeneous solutions of Eqs.~(\ref{3}) and~(\ref{3'}). For simplicity, we consider the case when the last term in Eq.~(\ref{3}) can be neglected, which is legitimate when the coupling constant~$\gamma$ or a small amplitude~$W$ is small (we will see that at temperatures~$T$ or doping level~$\mu$ near some critical values~$T_{N}$ and~$\mu_{N}$ the amplitude~$W$ is indeed small). In the zero-order approximation we obtain for~$\Delta(x)$
\begin{equation}
- \xi_{s}^{2} \Delta_{0}^{\prime \prime} + \Delta_{0} \big[ \Delta_{0}^{2} b_{s} - a_{s} \big] = 0\,.  \label{5}
\end{equation}
Equation~(\ref{5}) has the well-known nonuniform solution (see for example Ref.~\onlinecite{deGennes}),
\begin{equation}
\Delta_{0}(x) = \Delta_{\infty} \tanh (\kappa_{s} x) \,, \label{6}
\end{equation}
where ${\Delta_{\infty} = \sqrt{a_{s} / b_{s}}}$ and ${\kappa_{s}^{2} = a_{s} / 2 \xi_{s}^{2}}$. This equation describes, for instance, the behavior of~$\Delta(x)$ in a vicinity of S/N~interface at the superconductor side, where~N is a normal metal with a strong depairing. We consider this solution in an infinite superconductor.

Substituting this expression into Eq.~(\ref{3'}), we obtain an equation for
the amplitude of the CDW or SDW
\begin{equation}
\tilde{\xi}_{w}^{2} W^{\prime \prime} + W \big[ \mathcal{E} + \mathcal{U}_{w} \cosh^{-2}(\kappa_{s} x) \big] = g W^{3} \,, \label{7}
\end{equation}
where ${\mathcal{E} = a_{w} b_{s} - a_{s} \gamma}$, ${\mathcal{U}_{w} = a_{s} \gamma}$, ${g = b_{s} b_{w}}$, and ${\tilde{\xi}_{w}^{2} = \xi_{w}^{2} s_{3}}$. These quantities may be written in notations used for quasi-one-dimensional supercondcutors and Fe\nobreakdash-based pnictides as ${\mathcal{E} = \eta \big[ s_{3} (1 - \beta_{1}) - s_{2m} \big]}$, ${\mathcal{U}_{w} = \eta s_{2m}}$, ${g = s_{3} s_{3m}}$. Equation~\eqref{7} for spatial variation of the CDW amplitude~$W$ has a form of the well known Gross--Pitaevskii equation.\cite{Gross,Pitaev} Solutions of this equation can be written rather easily in limiting cases. We consider the simplest situation when the RHS of Eq.~(\ref{7}) is small, i.e., ${g W^{2} \ll \mathcal{U}_{w}}$.

We are interested in solutions with~$\Delta(x)$ given by Eq.~(\ref{6}) and~$W(x)$ decaying to zero at ${x \rightarrow \pm \infty}$. In particular, the solution for~$W(x)$ may have the form of a soliton. Such a state with a finite~$\Delta$ and zero~$W$ at infinity is stable if the conditions ${\partial^{2} \mathcal{F} / \partial \Delta^{2} > 0}$ and ${\partial^{2} \mathcal{F} / \partial W^{2} > 0}$ at ${\Delta = \Delta_{\infty} \neq 0}$ and ${W = 0}$ are satisfied. One can see that ${\partial^{2} \mathcal{F} / \partial \Delta^{2}|_{W=0} \sim b_{s}}$ is always positive and ${\partial^{2} \mathcal{F} / \partial W^{2}|_{W=0} \sim a_{w} - a_{s} \gamma / b_{s} \sim -\mathcal{E} / b_{s}}$ is positive if the quantity~$\mathcal{E}$ is negative. We will see that just at negative~$\mathcal{E}$, Eq.~(\ref{7}) has a solution in the form of a soliton.

In zero-order approximation we obtain for~$W_{0}$
\begin{equation}
\tilde{\xi}_{w}^{2} W_{0}^{\prime \prime} + W_{0} \big[ \mathcal{E} + \mathcal{U}_{w} \cosh^{-2}(\kappa_{s} x) \big] =0 \,. \label{8}
\end{equation}
This equation is integrable and its solutions~$\psi_{n}$ corresponding to a discrete spectrum of~$\mathcal{E}_{n}$ are expressed in terms of hypergeometric functions.\cite{LLquantMech} In our notations, the ``energy'' levels of discrete spectrum are given by\cite{LLquantMech}
\begin{equation}
\mathcal{E}_{n} = -a_{s} b_{s} \frac{\xi_{w}^{2}}{8 \xi_{s}^{2}} \Bigg[ -(1+2n) + \sqrt{1 + \frac{8 \gamma \xi_{s}^{2}}{\xi_{w}^{2} b_{s}}} \Bigg]^{2} \,, \label{9'}
\end{equation}
and their maximal number~$n_{\max }$ is determined by ${2 n_{\max} \leq \sqrt{1 + 8 \gamma \xi_{s}^{2} / \xi_{w}^{2} b_{s}} - 1}$. Note that in Fe\nobreakdash-based pnictides, ${\xi_{s} / \xi_{w} \simeq T_{w}/(T_{s} s_{3m}) = r / s_{3m}}$ in the ballistic case.

We expand the correction~$\delta W$ to the zero-order solution~$W_{0}$ in terms of the normalized eigenfunctions~$\psi_{n}$ of the operator ${\hat{\mathcal{L}} = -\tilde{\xi}_{w}^{2} \partial_{xx}^{2} - \mathcal{U}_{w} \cosh^{-2}(\kappa_{s}x)}$. These functions obey the equation
\begin{equation}
\hat{\mathcal{L}} \psi_{n} = \mathcal{E}_{n} \psi _{n} \,.  \label{11}
\end{equation}

Solutions of Eq.~(\ref{7}) can be written explicitly if the quantity ${\mathcal{E} = \mathcal{E}(\eta, \delta [\mu^{2}])}$ is close to a certain ``energy'' level~$\mathcal{E}_{n}$, say to~$\mathcal{E}_{N}$, such that ${\mathcal{E} \simeq \mathcal{E}_{N} = \mathcal{E}(\eta_{N}, \delta [\mu_{N}^{2}])}$ (in the language of the original electronic model, the ``temperature''~$\eta$ or doping~$\delta [\mu^{2}]$ should be chosen properly). We write Eq.~(\ref{7}) in the form
\begin{equation}
\hat{\mathcal{L}} W = \mathcal{E}_{N} W + R(W) \,, \label{12}
\end{equation}
with ${R = g W^{3} + (\mathcal{E} - \mathcal{E}_{N}) W}$ and represent~$W$ as ${W(x) = c_{N} \psi_{N}(x) + \delta W_{N}(x)}$, where ${\delta W_{N}(x) = \sum_{n}^{\prime} c_{N,n} \psi_{n}(x)}$, and the summation runs over all~$n$ except the term ${n = N}$. We substitute this~$W(x)$ into Eq.~(\ref{12}) and multiply this equation first by~$\psi_{N}$ and then by~$\psi_{n}$ with ${n \neq N}$, then integrating the obtained result each time over~$x$. Thus, taking into account the orthogonality of different eigenfunctions, we find the coefficients $c_{n}$
\begin{align}
c_{N}^{2} &= \frac{\mathcal{E} - \mathcal{E}_{N}}{g \langle\langle \psi_{N}^{4} \rangle\rangle} \,, \label{15} \\
c_{N,n} &= g c_{N}^{3} \frac{\langle\langle \psi_{N}^{3} \psi_{n} \rangle\rangle}{E_{n} - E_{N}} \qquad \text{with ${n \neq N}$} \,,  \label{15'}
\end{align}
where $\langle\langle f(x) \rangle\rangle = \int_{-\infty }^{\infty} \mathrm{d} x \, f(x)$, where the double angle brackets are used to distinguish this operation from the averaging over the angles introduced in the Appendix. Obviously, in Eq.~(\ref{15'}),~$\psi_{n}$ and~$\psi_{N}$ have to have same parity (both even or both odd).

The obtained expressions are valid provided the condition ${|\mathcal{E} - \mathcal{E}_{N}| \ll |\mathcal{E}_{n} - \mathcal{E}_{N}|}$ is satisfied. This condition means that if the ``temperature''~$\eta$ or doping~$\delta [\mu^{2}]$ is chosen in such a way that the quantity~$\mathcal{E}(\eta, \delta [\mu^{2}])$ is close to~$\mathcal{E}_{N}$, i.e., the difference on the LHS of this condition is smaller than the difference between any energy level~$\mathcal{E}_{n}$ and~$\mathcal{E}_{N}$, the spatial dependence of~$W(x)$ is given by the leading order while the second term,~$\delta W_N(x)$, gives a small correction. Since we assumed that the RHS of Eq.~(\ref{7}) is small compared to the term~$a_{s} \Delta$, the condition ${|\mathcal{E} - \mathcal{E}_{N}| \ll a_{s} b_{s} b_{w} / \gamma}$ should be also satisfied.

\begin{figure}[tbp]
\includegraphics[width=0.7\columnwidth]{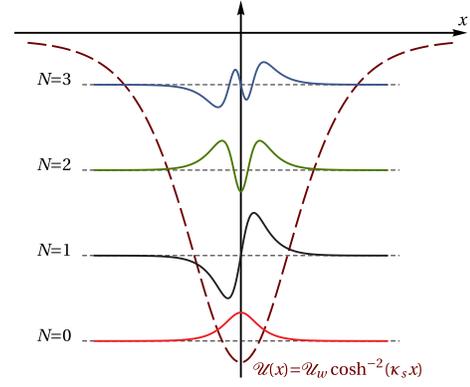}
\caption{(Color online.) Coordinate dependence of $W_0$~(red), $W_1$~(black), $W_2$~(green) and~$W_3$~(blue) near the corresponding ``energy'' levels~$\mathcal{E}_N$ for the case when the superconducting state is favored far from the defect at ${x = 0}$. Exactly at~$\mathcal{E}_N$, as follows from Eqs.~(\ref{15}) and~(\ref{15'}), ${W = 0}$. Note that in the opposite case when the CDW or the SDW state is more favorable at ${x \to \infty}$, one needs to make the exchange ${\Delta \leftrightarrow W}$ and, correspondingly, ${\mathcal{U}_{w} \leftrightarrow \mathcal{U}_{s}}$ and ${\kappa_{s} \leftrightarrow \kappa_{w}}$, and the shown curves will describe the dependence~$\Delta(x)$ while ${W = W_{\infty} \tanh (\kappa_{w} x)}$.}
\label{fig:W_n}
\end{figure}

The ground state is realized if at some ``temperature''~$\eta _{0}$ the quantity~$\mathcal{E}(\eta_{0})$ is close to~$\mathcal{E}_{0}(\eta_{0})$. In this case,~$W_{0}(x)$ has the form of a soliton. If $\mathcal{E}(\eta_{1})$ is close to~$\mathcal{E}_{1}$, the amplitude of the CDW is an odd function of~$x$. For the ground state, Eq.~(\ref{9'}) yields
\begin{equation}
\Bigg (A_{2} + \frac{\big\langle \delta [\mu^{2}] B_{2} \big\rangle }{\eta_{0}} \Bigg) = s_{3} \frac{s_{3m}}{8 r^{2}} \Bigg[ \sqrt{ 1 + \frac{8 r^{2} s_{2m}}{s_{3m} s_{3}}} - 1 \Bigg]^{2} \,,
\label{17}
\end{equation}
where ${A_{2} = s_{2m} - s_{3}(1 - \beta_{1})}$ and ${B_{2} = s_{3} \beta_{2}}$. At a given ${\delta [\mu^{2}] = \mu^{2} - \mu_{c}^{2}}$, that can be both positive and negative, this equation determines the ``temperature''~$\eta_{0}$ at which the solution of the Gross--Pitaevskii equation has a soliton-like solution~$W_{0}$. Similarly, setting ${n = 1}$ in Eq.~(\ref{9'}), one can find a ``temperature''~$\eta_{1}$ corresponding to the first excited state with an odd function of the OP~$W_{1}(x)$ etc. In Fig.~\ref{fig:W_n} we plot the spatial dependence of the CDW amplitude~$W(x)$ for ${n = 0}$,~$1$,~$2$ and~$3$. Note that if ${r^{2} s_{2m} / (s_{3m} s_{3}) < 1}$, only a single soliton-like solution exists.

\begin{figure}[tbp]
\includegraphics[width=0.5\columnwidth]{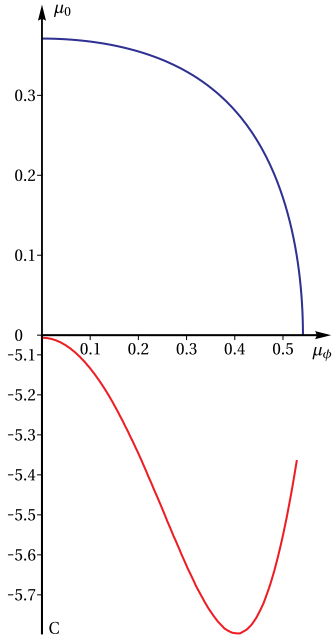}
\caption{(Color online.) The coefficient~$C$ in Eq.~(\ref{eq:eta_on_mu}) on~$\mu_c$ for ${r = 5.0}$. The critical doping~$\mu_c$ is calculated from~Eq.~(\ref{2}) and represents a line in the $(\mu_{\varphi}, \mu_0)$~plane, which is shown in the upper part of the figure. In the lower part, the value of~$C$ along the obtained $\mu_c$-line is presented as function of~$\mu_{\varphi}$ (inserting the corresponding value of~$\mu_0$). The coefficient~$C$ is negative, thus, as~${\eta > 0}$ due to conditions of realization of the superconducting state far from the topological defect, from Eq.~(\ref{eq:eta_on_mu}) it is seen that ${\delta[\mu^2] < 0}$.}
\label{fig:eta_on_mu}
\end{figure}

As an example, we calculate for the ground state the dependence of~$\eta_0$ on~$\delta[\mu^2]$. It follows from Eq.~(\ref{17}) that, assuming~$\delta[\mu^2]$ independent on~$\varphi$,
\begin{equation}
\eta_0 = C \cdot \delta[\mu^2] \,,
\label{eq:eta_on_mu}
\end{equation}
where the coefficient~$C$ is given by
\begin{equation}
C = \langle B_2 \rangle \Bigg[ \frac{s_3 s_{3m}}{8r^2} \Bigg( \sqrt{1 + \frac{8 r^2 s_{2m}}{s_3 s_{3m}} - 1} \Bigg)^2 - A_2 \Bigg]^{-1} \,.
\end{equation}
The coefficient~$C$ depends on~$\mu_c$ defined by~Eq.~(\ref{2}). It is negative and, thus, since ${\eta > 0}$, $\delta[\mu^2]$ should also be negative. We plot the dependence of~$C$ on~$\mu_c$ in Fig.~\ref{fig:eta_on_mu}. More precisely, the critical doping~$\mu_c$ is calculated from~Eq.~(\ref{2}) and represents a line in the $(\mu_{\varphi}, \mu_0)$~plane (the upper part of~Fig.~\ref{fig:eta_on_mu}). Projecting this line onto the $\mu_{\varphi}$~axis and inserting the corresponding values of~$\mu_0$ we obtain the plot of~${C = C(\mu_c) \equiv C(\mu_0, \mu_{\varphi})}$ presented in the lower part of~Fig.~\ref{fig:eta_on_mu}.

Consider the temperature interval where the soliton-like solution for~$W(x)$ exists. As follows from Eq.~(\ref{15}), the difference ${\mathcal{E} - \mathcal{E}_{0}}$ must be positive if the constant ${g = b_{s} b_{w}}$ is positive. This implies that the difference ${\mathcal{E} - \mathcal{E}_{0} = A_{2} (T - T_{0})}$ has to be positive as well (at ${A_{2} > 0}$). Therefore, at ${T < T_{0}}$, no~$W$ appears at the topological defect, but at~${T > T_{0}}$, a soliton-like solution for~$W(x)$ arises with the amplitude ${W(0) \sim \sqrt{(T - T_{0})}}$. On the other hand, as follows from Eq.~(\ref{17}), the temperature~$T_{0}$ is less than the temperature ${T_{2} \equiv c_{\mu} B_{2} / A_{2}}$, where ${c_{\mu} = -\delta [\mu^{2}]}$. This means that the soliton-like solution for~$W$ as well as solutions corresponding to excited states exist in the interval
\begin{equation}
T_{0} < T < T_{2} \,. \label{17a}
\end{equation}

The solutions found above are valid if the free energy~$\mathcal{F}_{s}$ of the superconducting state at ${x \rightarrow \pm \infty}$ is lower than~$\mathcal{F}_{w}$ for a state with ${W \neq 0}$. This is possible if ${\mathcal{F}_{s} - \mathcal{F}_{w} \sim (\Delta^{4} - W^{4}) \sim \big[ a_{s}(\eta) / b_{s} - a_{w}(\eta) / b_{w} \big] < 0}$. This condition determines a temperature interval in which our considerations are valid.

If the difference ${\mathcal{F}_{s} - \mathcal{F}_{w}}$ is positive, then the same procedure of finding solutions of G\nobreakdash--L~equations can be repeated with an exchange ${\Delta \leftrightarrow W}$ adapting correspondingly~$\mathcal{E}_{n}$ and other quantities. In particular, ${W = W_{\infty} \tanh (\kappa_{w} x)}$, with ${W_{\infty} = a_{w} / b_{w}}$ and ${\kappa_{w} = \sqrt{2} a_{w} /\xi_{w}}$, and the superconducting OP~$\Delta(x)$ is expressed in terms of hypergeometric functions, i.e., it is localized at ${x = 0}$. Consider, for example, an N/S~system where in the superconductor~S there exists not only the superconducting OP~$\Delta$, but also a density wave~$W$, and~N is a normal metal with a strong suppression of~$W$ (for example, with a strong interband impurity scattering which suppresses the OP~$W$~\cite{Vavilov11,*Vavilov11a,*Vavilov11b,Syzranov14}). Then, at the S\nobreakdash-side, the dependence~$W(x)$ is determined by the above written expression, and at a certain temperature which may be even higher than~$T_{s}$, at the N/S~interface superconductivity may arise spreading over a distance~${\sim \xi_{s}}$ from the interface.

Note that the found nonhomogeneous solutions for~$\Delta(x)$ and~$W(x)$ are energetically favorable in comparison with uniform solutions,~$\Delta_{\infty}$ and~$W_{\infty}$, provided the energy loss [due to the gradient of~$\Delta(x)$] ${\delta \mathcal{F}_{s} \sim \int \mathrm{d} x \, \Delta_{\infty}^{4} [ 1 - \tanh^{4} (\kappa_{s} x) ] \sim \Delta_{\infty}^{4} \xi_{s} / \sqrt{a_{s}}}$ is less than the energy gain~$\delta \mathcal{F}_{w}$ (due to the appearance of~$W$) ${\delta \mathcal{F}_{w} \sim \int \mathrm{d} x \, W^{4}(x) \sim W_{\infty}^{4} \xi_{w}}$. This cannot occur in the considered case of small~$W$. However, in heterostructures, like an N/S~system, the solution Eq.~(\ref{6}) (at ${x > 0}$) is dictated by a boundary condition in case of strong depairing in the N~metal and, therefore, there is no energy loss in the superconducting part of the free energy. Thus, the considered states may be realized in heterostructures. The case of uniform superconductors with a not small OP~$W$ requires a separate consideration.

\section{Conclusion}

On the basis of Ginzburg--Landau equations we studied a possibility of nonhomogeneous states in systems with two OPs. Materials, where the superconducting OP~$\Delta$ and the OP~$W$ related to a CDW (or an SDW) may exist, belong to this class of systems. In the situation
when the superconducting state is more favorable, the Ginzburg--Landau equations have nonhomogeneous solutions which describe~$\Delta(x)$ in the form of a topological defect, Eq.~(\ref{6}), and~$W(x)$---in the form of a function localized near the center of the defect, ${x = 0}$. The form of~$W(x)$ is described by the Gross--Pitaevskii equation and depends essentially on the proximity of the function~$\mathcal{E}(\eta, \delta [\mu^{2}])$ to the eigenvalues~$\mathcal{E}_{N}$ of the linearized Gross--Pitaevskii equation. If~${\mathcal{E}(\eta, \delta [\mu^{2}]) = \mathcal{E}_{N}}$ at some temperature~${T_{N} = (1 - \eta_{N}) T_{s}}$ and doping~$\delta [\mu_{N}^{2}]$, then the amplitude of the function~$W(x)$ turns to zero and increases as ${W \sim \sqrt{|\mathcal{E}_{N} - \mathcal{E}(\eta, \delta [\mu^{2}])|}}$ when~$\eta$ or~$\delta [\mu^{2}]$ deviate from~$\eta_{N}$ and~$\delta [\mu_{N}^{2}]$. At a given temperature~$T$ in the interval Eq.~(\ref{17a}), there are, generally speaking, several solutions for~$W(x)$. The most stable one  is the solution which corresponds to the ground state (soliton-like solution). Therefore, in the equilibrium case one can observe only this solution for~$W(x)$. Other solutions may affect the response of the system to the influence of
fluctuations or of external perturbations.

On the other hand, if the state with ${W \neq 0}$ and ${\Delta = 0}$ corresponds to a minimum of the free energy, then nonhomogeneous solutions are possible with~$W(x)$ determined by Eq.~(\ref{6}) (correspondingly adapted as ${\Delta \rightarrow W}$, ${\xi_{s} \rightarrow \xi_{w}}$) and~$\Delta(x)$ is localized near the point ${x = 0}$. In principle, such solutions may arise in the bulk (especially near some defects) and in heterostructures of type N/N$_{s,w}$, where~N$_{s,w}$ is a material under consideration in which~$\Delta$ and/or~$W$ may exist, and~N is a material with a strong depairing towards the OPs~$\Delta$ and~$W$. For example, in an~N/N$_{w}$ heterostructure, the OP in the vicinity of the interface has inevitably the form of Eq.~(\ref{6}) and a localized~${\Delta(x) \neq 0}$ arises at the interface. In this case, one deals with a localized interfacial superconductivity. This type of superconductivity has been studied very actively in recent years and has been observed in different materials including cuprates and Fe\nobreakdash-based pnictides (see recent papers Refs.~\onlinecite{Richter_et_al_2013,Reich_2013,Wu_et_al_2013} and references therein). Several proposals have been made to explain this phenomenon, but most experimental observations remain unexplained. The mechanism considered here may be responsible for interfacial superconductivity in systems with two OPs, but applicability of this mechanism to real materials deserves a separate consideration.

\acknowledgments

We appreciate the financial support from the DFG via the Projekt~EF~11/8\nobreakdash-1; K.~B.~E.~gratefully acknowledges the financial support of the Ministry of Education and Science of the Russian Federation in the framework of Increase Competitiveness Program of  NUST~``MISiS'' (Nr.~K2-2014-015).

\appendix

\section{Coefficients in the Ginzburg--Landau equations}

The free energy has the form (see also Refs.~\onlinecite{Chubukov10,*Chubukov10a,Schmalian10})
\begin{equation}
\Phi (\Delta, W, \mu) = -(2 \pi T) \sum_{\omega = 0}^{E_{\text{m}}} \Re(P) + \frac{\Delta^{2}}{2 \lambda_{\text{sc}}} + \frac{W^{2}}{2 \lambda_{\text{dw}}} \,, \label{A1}
\end{equation}
where ${P = \sqrt{(\varsigma_{\text{sc} \omega} + \mathrm{i} \mu)^{2} + W^{2}}}$, ${\varsigma_{\text{sc} \omega} = \sqrt{\omega^{2} + \Delta^{2}}}$ and~$\omega$ is the Matsubara frequency with a cut-off~$E_{\text{m}}$; $\lambda_{\text{sc}}$ and $\lambda_{\text{dw}}$ are the interaction constants of the superconductivity and spin- or charge-density wave, respectively. Expanding this expression in~$\Delta$ and~$W$ and performing variation with respect to these variables, we come to Eqs.~(\ref{1}) and~(\ref{1'}) with the coefficients defined as
\begin{align}
s_3 &= \sum_{n=0}^{\infty}
(2 n + 1)^{-3} \,, \\
s_{1m} &= \sum_{n=0}^{\infty}
(2 n + 1)^{-1} \big[ (2 n + 1)^2 t^2 + m^2 \big]^{-1} \,, \\
s_{2m} &= \sum_{n=0}^{\infty}
\big\langle \big[ (2 n + 1)^2 - m^2 \big] (2 n + 1)^{-1} \big[ (2 n + 1)^2 + m^2 \big]^{-2} \big\rangle \,, \\
s_{3m} &= \sum_{n=0}^{\infty}
\big\langle (2 n + 1) \big[ (2 n + 1)^2 - 3 m^2 \big] \big[ (2 n + 1)^2 + m^2 \big]^{-3} \big\rangle \,, \\
\beta_1 &= \sum_{n=0}^{\infty}
\big\langle 4 m^2 (2 n + 1) \big[ (2 n + 1)^2 + m^2 \big]^{-2} \big\rangle \,, \\
\beta_2 &= \sum_{n=0}^{\infty}
2 (2 n + 1)^{-1} \big[ (2 n + 1)^2 + m^2 \big]^{-1} \,,
\end{align}
where ${t = T/T_{s}}$ and the angle brackets denote the angle averaging (in Fe\nobreakdash-based pnictides) or integration along the sheets of the Fermi surfaces in quasi-one-dimensional superconductors.


%

\end{document}